\documentclass[useAMS,manuscript]{emulateapj}

\usepackage{verbatim}
\usepackage{natbib}
\usepackage{amsmath}
\usepackage{amsbsy}
\usepackage{lscape}
\usepackage{epstopdf}

\slugcomment{Submitted to ApJ}
\shorttitle{Barnard 68}
\shortauthors{Burkert \& Alves}

\begin{document}

\title{The Inevitable Future of the Starless Core Barnard 68}

\author{Andreas Burkert\altaffilmark{1} and Jo\~{a}o Alves\altaffilmark{2}}

\altaffiltext{1}{University Observatory Munich, Scheinerstrasse 1, D-81679 Munich}
\altaffiltext{2}{Calar Alto Observatory, C. Jes\'us Durb\'an Rem\'on, 2-2, E-4004
Almeria, Spain}

\email{burkert@usm.uni-muenchen.de, jalves@caha.es}

\newcommand\msun{\rm M_{\odot}}
\newcommand\lsun{\rm L_{\odot}}
\newcommand\msunyr{\rm M_{\odot}\,yr^{-1}}
\newcommand\be{\begin{equation}}
\newcommand\en{\end{equation}}
\newcommand\cm{\rm cm}
\newcommand\kms{\rm{\, km \, s^{-1}}}
\newcommand\K{\rm K}
\newcommand\etal{{\rm et al}.\ }
\newcommand\sd{\partial}

\begin{abstract}
Dense, small molecular cloud cores have been identified as the
direct progenitors of stars. One of the best studied examples is
Barnard 68 which is considered a prototype stable, spherical gas
core, confined by a diffuse high-pressure environment.
Observations of its radial density
structure however indicate that Barnard 68 should be gravitationally
unstable and collapsing which appears to be inconsistent with its
inferred long lifetime and stability. We argue that Barnard 68
is currently experiencing a fatal collision with another small core
which will lead to gravitational collapse. Despite the fact that
this system is still in an early phase of interaction, our numerical
simulations imply that the future gravitational collapse is already
detectable in the outer surface density structure of the globule
which mimicks the profile of a gravitationally unstable Bonnor-Ebert
sphere. Within the next $ 2 \times 10^5$ years Barnard 68 will
condense into a low-mass solar-type star(s), formed in isolation,
and surrounded by diffuse, hot interstellar gas. As witnessed in
situ for Barnard 68, core mergers might in general play an important
role in triggering star formation and shaping the molecular core
mass distribution and by that also the stellar initial mass
function.
\end{abstract}


\section{Introduction}

Barnard 68 is considered an excellent test case and the prototype of a
dense molecular cloud core (Alves et al. 2001b). Because of its small distance
($\sim$ 125 pc), this so called Bok globule /Bok \& Reilly 1947) with a mass of M=2.1 $M_{\odot}$,
contained within a region of R = 12,500 AU has
been observed with unprecedented accuracy (Alves et al. 2001b; Lada et al. 2003;
Bergin et al. 2006; Redman et al. 2006; Maret et al. 2007).  Deep
near-infrared dust extinction measurements of starlight toward
individual background stars observed through the cloud provided a
detailed 2-dimensional map of its projected surface density
distribution (Fig. 1) from which a
high-resolution density profile was derived over the entire extent of
the system. The striking agreement of the inferred density structure
with the theoretical solution of an isothermal, pressure confined,
hydrostatic gas sphere (so called Bonnor-Ebert sphere) was interpreted
as a signature that the globule is old, thermally supported and stable,
with the pressure gradient balancing the gravitational force. 

This conclusion has received additional support from molecular line
observations (Lada et al. 2003) that show complex profile shapes which can be
interpreted as signatures of stable oscillations (Redman et al. 2006; Broderick et al. 2007) 
with subsonic velocities of order $V \approx$ 0.04 km/s which is 20\% of the
isothermal sound speed $c_s \approx$ 0.2 km/s. The age of Barnard 68
should therefore be larger than one dynamical oscillation timescale
$\tau_{dyn}$ = 2R/V = 3 $\times 10^6$ yrs which is long compared to its
gravitational collapse timescale $\tau_{coll}=(3 \pi/32 G \rho)^{1/2}
= 0.17 \times 10^6$ years, where $\rho=1.5 \times 10^{-19}$ g cm$^{-3}$ is
the average mass density. The observed hydrostatic profile and the
inferred large age are strong arguments for stability.

Other observations however are in conflict with this
conclusion. The best fitting hydrostatic model leads to the
conclusion that Barnard 68 should be gravitationally
unstable (Alves et al. 2001b).  Pressure confined, self-gravitating gas spheres with radii $R$
and isothermal sound speeds $c_s$ have
self-similar density distributions (see Appendix) that are
characterized by the dimensionless parameter

\begin{equation}
\xi_{max} = \frac{R}{c_s} \sqrt{4 \pi G \rho_c}
\end{equation}

\noindent where $\rho_c$ is the central density.  Although all values
of $\xi_{max} > 0$ represent equilibrium solutions where the
gravitational force is exactly balanced by pressure forces, a
stability analyses (Bonnor 1956) shows that small perturbations should lead to
gravitational collapse in cores with $\xi_{max} > 6.5$.  Barnard 68's
surface density distribution is characterised by $\xi_{max} = 6.9
\pm 0.2$. It should collapse on a free-fall timescale which
is much shorter than its oscillation timescale. In addition, the
question arises how this globule could ever achieve an unstable
equilibrium state in the first place.

Dense molecular cores like Barnard 68 are the last well characterized
configuration of interstellar gas before gravitational collapse
towards star formation. They have masses from a tenth to tenths of
solar masses and typical sizes from a tenth to a few tenths of a
parsec (Alves et al. 2007).  They are found in isolation, known then as Bok
globules (Bok 1948), or embedded in lower density molecular cloud
complexes. Because of their relatively simple shapes they have long
been recognized as important laboratories to the process of star
formation. Barnard 68 is one such dense core.  Despite being isolated
and surrounded by warm and diffuse gas it is part of the Pipe Nebula
complex (Alves et al. 2007; Lombardi et al. 2006; Lada et al. 2008) 
which consists of an ensemble of about 200
cores within a region of order 10 pc.  Many cores
appear distorted and asymmetric. This could be a result of their
interaction with the highly turbulent diffuse gas environment that leads
to non-linear and non-radial oscillations (Broderick et al., 2007, 2008). 
Another possibility which we will investigate here are collisions between cores.
We propose that Barnard 68 is
currently experiencing such a fatal collision that triggers
gravitational collapse. Its peculiar and seemingly contradictory
properties are early signatures of this process that cannot be
understood if the globule is treated as an isolated and stable
Bonnor-Ebert sphere. Observations
show that the distribution of molecular core masses is surprisingly similar
to the stellar initial mass function (Alves et al. 2007). This indicates that
the masses of stars originate directly from the processes
that shape the molecular core mass function.
Barnard 68 demonstrates that core-core collisions could play
an important role in this process.

\section{Evidence For a Collision}

The surface density map of Barnard 68 (upper panel of Fig. 1) clearly shows a
southeastern prominence which appears to be a separate smaller globule
(the bullet) that is currently colliding with its larger companion
(the main cloud). From the bullet's surface density distribution we
infer a gas mass of 0.2 $M_{\odot}$ masses which is 10\% the mass of the
main cloud. Note that the bullet appears tidally elongated, probably
along its orbit that is oriented perpendicular to the line of
sight. Such a distortion is expected if the mutual gravitational
attraction could affect the bullet during its approach, that is if its
encounter velocity is of order or smaller than $(GM/R)^{1/2} \approx$
0.4 km/s.

\begin{figure}[ht]
\begin{center}
\includegraphics[width=0.50\textwidth]{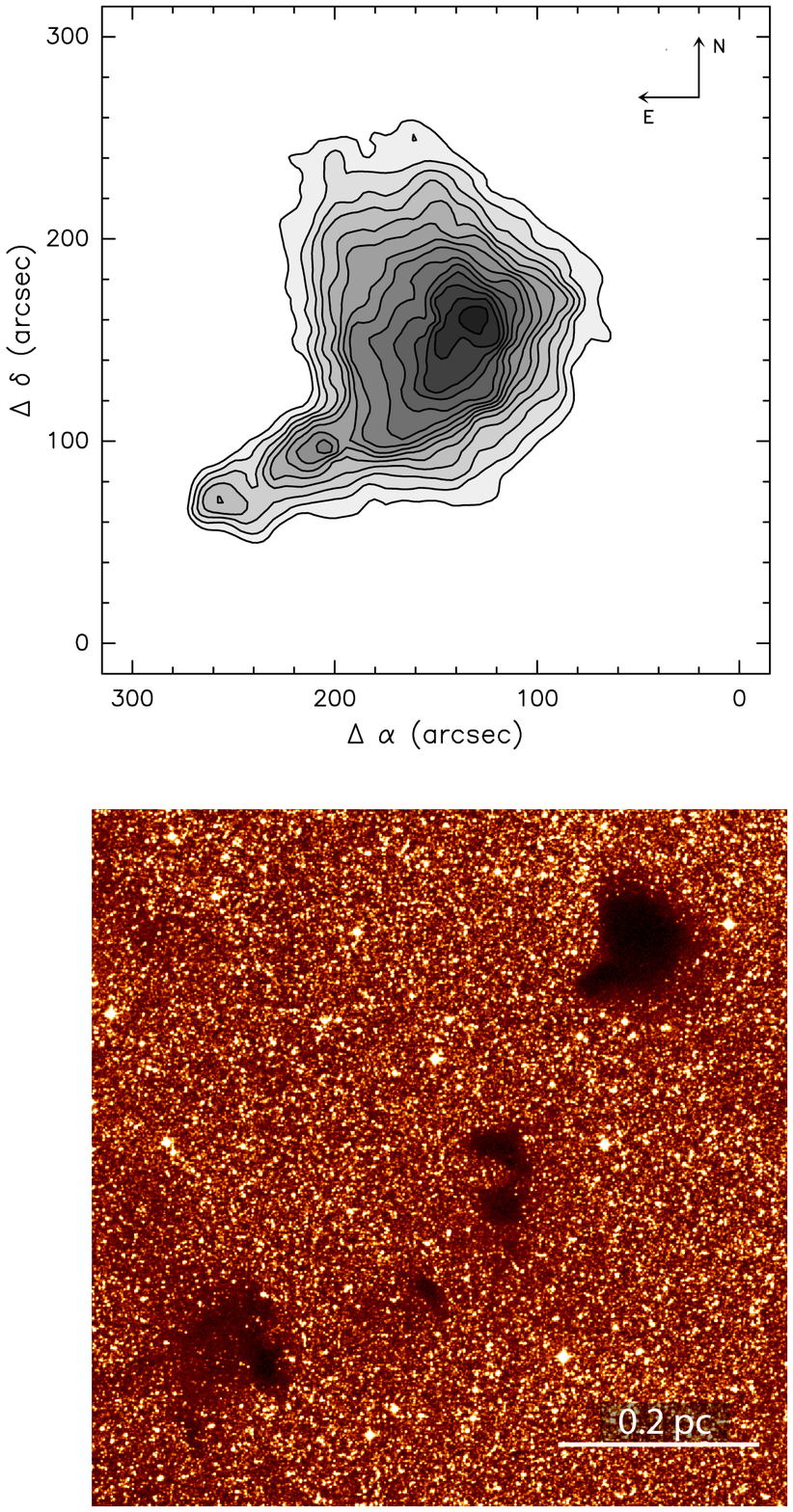}
\end{center}
\caption{
\label{fig1}
The upper panel shows the dust column density map of Barnard 68
(Alves et al. 2001a). About a
1000 measurements of line of sight extinction towards background
stars were used to construct this map. Contours start at 2 magnitudes
of optical extinction (A$_V$), and increase in steps of 2
magnitudes. The peak column density through the center of the cloud
is about 30 magnitudes of visual extinction. 
The optical image (0.68$\mu$m) of Barnard 68 (upper
right) and its immediate surroundings to the South-West are
shown in the lower panel.  These
clouds are part of the larger (10 pc) Pipe Nebula complex. Image
from the Digitized Sky Survey.}
\end{figure}

One might argue that the apparent collision is a projection effect and that B68 and the bullet are 
two isolated equilibrium clouds. There is however strong evidence for the merger
scenario. First of all, the bullet clearly shows substructure (Fig. 1). It consists of
two embedded components which indicates that the bullet itself is part of a merger.
If the bullet would be isolated, pressure confined and as long-lived as B68
it would not have two density peaks. This indicates that mergers occur in the Pipe
nebula. If the two small substructures in the bullet formed by a merger despite their small
geometrical cross section, it is even more likely that B68 can capture a clump due to
its large cross section and larger gravitational force.
In addition, the lower panel of figure 1 shows that B68, 
the two substructures in the bullet and the other clumps
in the vicinity of B68 are not randomly oriented but aligned like pearls on a string.
Numerical simulations (e.g. 
Burkert \& Bodenheimer 1993, 1996; Truelove et al. 1998; Burkert \& Hartmann 2004)
of molecular cloud collapse and star formation demonstrate
that such a situation can arise naturally when elongated gas clouds and sheets 
become gravitationally unstable and collapse, forming a dense filament of gas that will
fragments if the collapse perpendicular to the filament is stopped temporarily, e.g. due to heating
by the release of gravitational energy.  In this case, the simulations show that
the fragments move along the filament and merge.

\section{Numerical Simulations of Colliding Bonnor-Ebert Spheres}

\begin{figure*}[ht]
\begin{center}
\includegraphics[width=0.80\textwidth]{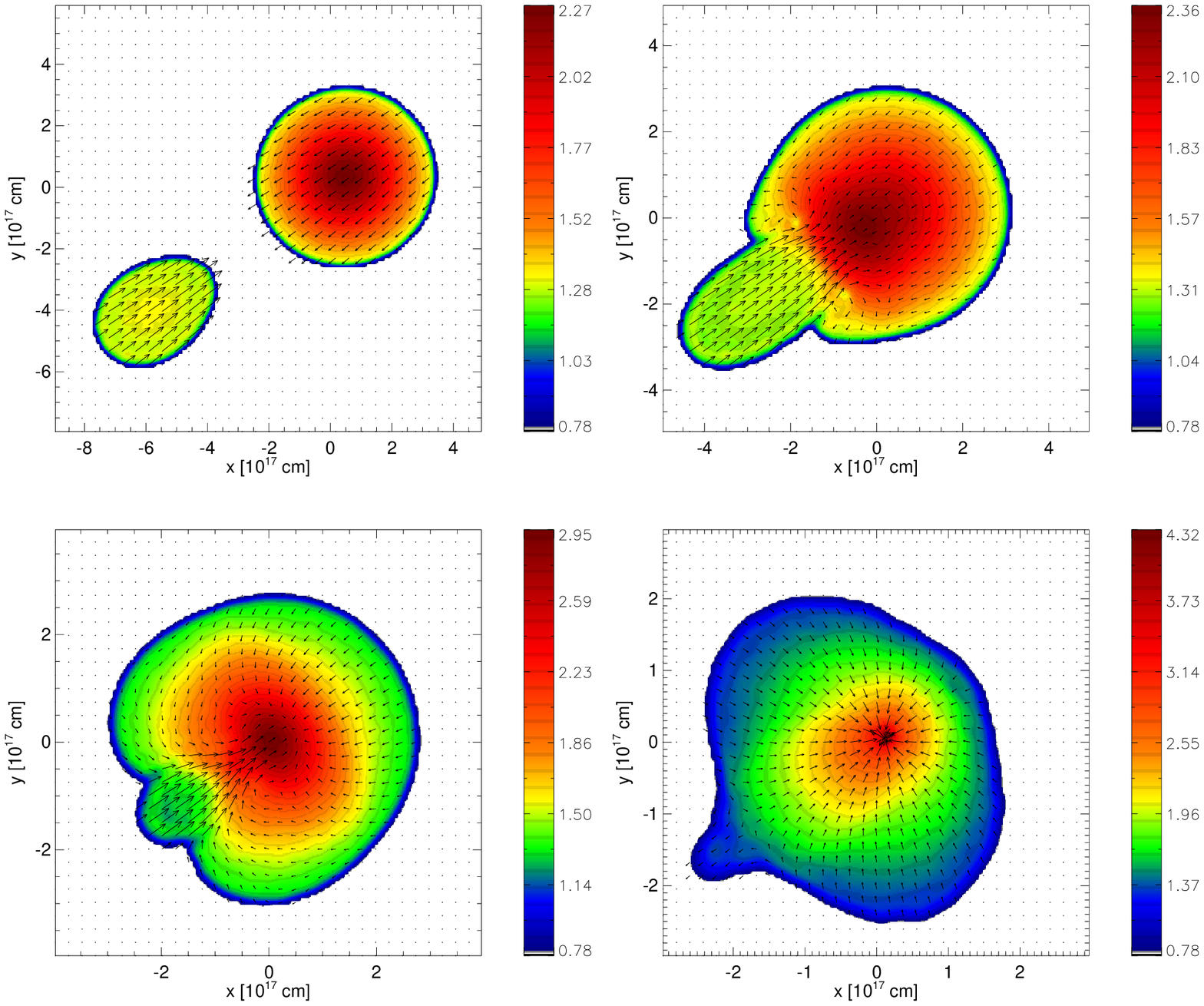}
\end{center}
\caption{
\label{fig2}
  Snapshots of our standard model, a merger
  simulation of a 0.21 solar mass bullet with a 2.1 solar mass main
  cloud that is characterised initially by a dimensionless parameter
  $\xi_{max}$ = 6. The panels show 2-dimensional cuts through the equatorial plane
 of the 3-dimensional SPH simulation.
 Colors indicate the logarithm of the gas density
  in units of $4 \times 10^{-15} g/cm^3$.  The arrows show the
  velocity field. The upper left diagram shows an early phase before
  the collision, 0.9 million years after the start of the
  simulation. The maximum velocities are 0.16 km/s. The bullet is
  already tidally elongated.  The upper right diagram represents the
  present state of Barnard 68, 1.7 million years after the start of
  the simulation. The
  lower left and lower right panels show the system 1.95 million years
  and 2.1 million years after the start of the simulations. The
  maximum velocities are 0.46 km/s and 4 km/s, respectively.  As soon
  has the bullet has merged with the main cloud, the whole system begins to
  collapse with large infall velocities and the formation of a
  high-density central core.}
\end{figure*}

We studied the collision of Barnard 68 with a small core numerically.
The isothermal hydrodynamical equations of merging gaseous globules
were integrated using the SPH algorithm (Wetzstein et al. 2008) with 40000
SPH particles for the main cloud and 4000 particles for the bullet.
The confining effect of the external pressure $P_{ext}$ was taken into
account by modifying the equation of state: $P = \rho c_s^2 -
P_{ext}$.

Two equilibrium cores with masses of 2.1 and 0.21 solar masses,
respectively, were generated, adopting an isothermal sound speed of
$c_s = 0.2$ km/s and no initial spin. In order to generate initial
conditions for different values of $\xi_{max}$, we determined the
corresponding dimensionless mass m (see Appendix) and from that calculated
the required external pressure 

\begin{equation}
P_{ext} = m^2 \frac{c_s^8}{G^3 M^2}.
\end{equation}

\noindent To guarantee hydrostatic equilibrium, the cores were allowed
to relax in isolation for 20 dynamical timescales. At the end of this
phase their surface density profiles followed the expected
Bonnor-Ebert solution with high accuracy. We took a main cloud with 
$\xi_{max}=6$, corresponding to a dimensionless mass of $m_{B68} = 1.179$
and an external pressure of $P_{ext}=6.8 \times 10^{-12}$ dyn cm$^{-2}$.
The bullet is embedded in the same pressure environment. Therefore its dimensionless
mass is $m_{bull} = 0.1 \times m_{B68} = 0.118$, corresponding to $\xi_{max}=1.16$. For the standard model the
bullet was placed at rest at a distance d=0.3 pc from the main
cloud.  The mutual gravitational force accelerated the bullet towards
the main cloud, leading to a collision after 1.7 million years.  

Fig. 2 shows the evolution  of our standard model.
The bullet is accelerated towards the
main cloud by their mutual gravitational attraction. After $1.7 \times 10^6$
yrs the bullet collides with the main cloud with a relative velocity
of 0.37 km/s which is supersonic.  As seen in the upper right diagram of Fig. 2, 
the bullet is
now clearly tidally elongated and its momentum is large enough for its
gas to penetrate deeply into the main cloud, all the way into its
central region. An interesting side effect of this process is that CO-rich
gas from the bullet and the outer parts of the main cloud is funneled into
the CO-depleted core region of the main cloud which should increase the core's CO abundance.
Theoretical models of detailed molecular line observations indicate
depletion of CO in the inner region of
Barnard 68 as a result of molecule freeze-out onto grain surfaces 
(Bergin et al. 2006; Maret et al. 2007). The
models predict chemical timescales of only 1-3 $\times 10^5$ yrs to reach
the observed degree of CO depletion which is an order of magnitude shorter
than the expected lifetime of the gas clump. For ages of order
a million years freeze-out should however be more efficient
and the C$^{18}$O(1-0) line intensities should be smaller than measured.
The merger might increase the central CO abundance enough in order to solve
this problem.
Note, that these conclusions are still a matter of debate due to the uncertainties in the 
chemical network and the treatment of radiation transport as well as due to uncertainties in
understanding the formation and past evolution of the globule.
The affect of the current merger on the internal chemical 
structure of Barnard 68 will be studied in details in a subsequent paper.

\begin{figure}[ht]
\begin{center}
\includegraphics[width=0.50\textwidth]{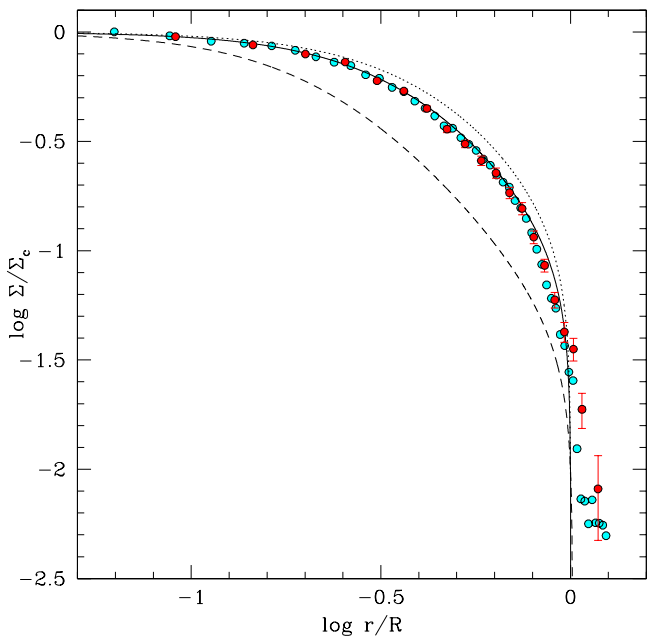}
\end{center}
\caption{
\label{fig3}
The cyan points show the surface density
  structure $\Sigma (r)$ of the main cloud at the onset of merging
  with the bullet (see the upper right panel of Fig. 2).  Here $\Sigma$ is normalized to the
  central value $\Sigma_c \equiv \Sigma (r=0)$ The radius r is
  normalized to the maximum radius R (the theoretical edge) of the
  main cloud if it would follow a perfect Bonnor-Ebert profile. Red
  points with error bars show the observed structure of Barnard 68 as
  determined from near IR dust extinction measurements. The dotted,
  solid and dashed lines show theoretical Bonnor-Ebert spheres with
  $\xi_{max}$ = 6, 7 and 10, respectively. The numerical simulations are
  in excellent agreement with the observations. The low surface
  density points beyond the theoretically expected edge, i.e. for $r >
  R$ correspond to perturbed gas at the edge of the main cloud and gas
  from the infalling bullet.}
\end{figure}

The observed bullet, despite having the same
mass as in the simulation, is more elongated and less spherical.
This is a result of the fact that the bullet itself has experienced
a merger recently. In our standard simulation we did not take this into account.
We have however verified through
test simulations that the physics of the encounter
and the subsequent evolution of the system
does not depend sensitively on the adopted initially elongation of the infalling
substructure or whether two almost attached substructures, each with 50\% the bullet's mass,
merge with the main cloud.

The collision disturbs the main cloud, leading to characteristic
features that can be compared with observations. Fig. 3 shows that the
surface density distribution (cyan points) of the core can still be
described very well by a Bonnor-Ebert fit (solid line).  However,
compared to the initial distribution with $\xi_{max} = 6$ (dotted
line), its profile is now characterized by a larger value of
$\xi_{max} = 7 \pm 0.2$ (solid line) which is in the unstable
regime. The structure of the simulated globule is in excellent
agreement with the observations (red points in Fig. 3). Note that
there is gas beyond the theoretical Bonnor-Ebert edge R of the main
cloud which mainly belongs to the infalling bullet.

Is there evidence for the collision in line-profile measurements of
the line-of-sight velocity distribution? 
Adopting a projection angle of 90 degrees, that is a line-of-sight
precisely perpendicular to the orbital plane, our simulations 
show that for an initially quiescent main
cloud without strong intrinsic oscillations, the gravitational
interaction with the bullet triggers a characteristic line-of-sight
velocity field that shows infall of the outer envelope of the main
cloud with a velocity of order 10 m/s while the ram pressure of the
merger leads to an outflow of gas from the center with a velocity of
up to 5 m/s. Indeed, gas outflow from the center and infall of the
outer envelope has been detected (Maret et al. 2007). However the measured velocities
are a factor of 4 larger than predicted theoretically. This might
indicate that the observed velocity structure is still dominated by
the expected natural stable oscillations (Redman et al. 2006; Broderick et al. 2007) 
of Barnard 68, prior to the
encounter. An impact exactly perpendicular to the line-of-sight is 
however unlikely. We find that the velocity distribution along the
line-of-sight depends on the adopted projection angle.  For
projection angles of 60-80 degrees, the line-of-sight
velocity profiles show mean outer infall motions of 40 m/s as they now
contain the signature of the fast encounter, in better agreement with the
observations. 

B68's observed
velocity field (Lada et al. 2003) shows an asymmetric distribution with
coherent infall with a velocity of 120 m/s in the southeastern part 
corresponding to the interaction region of the two clumps. If this
is interpreted as the line-of-sight velocity part of the bullet and assuming
an impact velocity of 0.4 km/s, the inclination angle would be 73 degrees.
In general, for projection angles of 60 degrees or larger (probability of 50\%)
the line-of-sight velocity distribution of the merger simulations is consistent with
a combination of intrinsic oscillations of B68 and gas flows
generated as a result of the merger. 

\section{Collision-Triggered Collapse of Barnard 68}

\begin{figure}[ht]
\begin{center}
\includegraphics[width=0.45\textwidth]{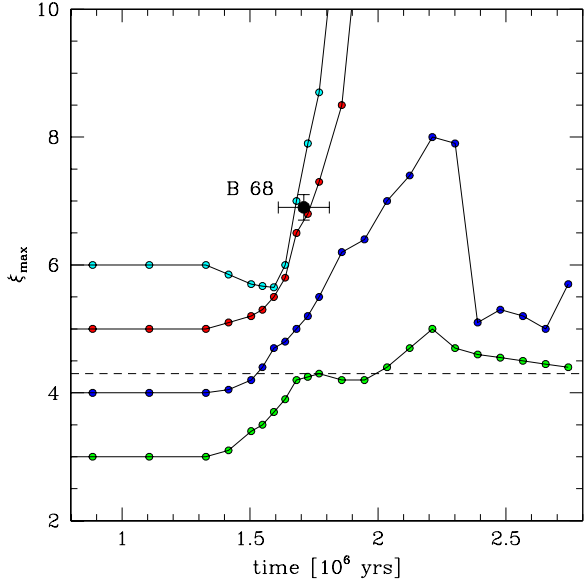}
\end{center}
\caption{
\label{fig4}
Evolution of the dimensionless parameter
  $\xi_{max}$ of the main cloud, adopting different initial values for
  $\xi_{max}(t=0)$. The main cloud experiences a collision with a
  bullet at t=1.3 million years.  The dashed curve divides the diagram
  into two regions.  For main clouds with $\xi_{max}(t=0) < 4.3$, the
  dimensionless mass m after the merger is still smaller than the
  critical mass for gravitational collapse ($m_{crit} < 1.15$). They go
  through a violent relaxation phase and then settle into a new stable
  Bonnor-Ebert state.  Objects that start initially above this line
  cannot achieve stability after the merger and therefore
  collapse. The green and blue points show two stable merger
  simulations that started with main clouds of $\xi_{max}$ = 3 and 4,
  respectively.  The red and cyan points correspond to unstable
  mergers with initial values of $\xi_{max}$ = 5 and 6. The black
  point with error bars shows the observed evolutionary state of
  Barnard 68 which is in very good agreement with an unstable merger
  that leads to gravitational collapse.}
\end{figure}

As shown in the two lower panels of Fig. 2, the main cloud collapses as soon
as the bullet has completely merged. This is not surprising as the dimensionless
mass of the total system $m_{tot}=m_{B68}+m_{bull}=1.3$ is larger that the critical
Jeans limit $m_{Jeans}=1.18$ for a stable Bonnor-Ebert sphere (see Appendix).
An initial stability parameter of $\xi_{max} = 6$ for B68 however
appears rather fine tuned. 
What is the range of $\xi_{max}$ values that is in agreement with
the observations? And could Barnard 68 survive the encounter as a stable cloud?

Fig. 4 shows
the time evolution of the dimensionless parameter $\xi_{max}$ for four simulations with
different values of the confining external pressure that correspond to
different initial values of $\xi_{max}$. The mass of the main cloud and 
the bullet were the same as in the standard model.
Theoretically, main clouds
with initial values of $\xi_{max} \leq 4.3$ (dashed line), corresponding to
the dimensionless mass $m \leq 1.07$ could
survive a merger with a bullet of 10\% their mass as $m_{tot} < m_{Jeans}$.
Indeed we find that for initial values of the main cloud
of $\xi_{max}$ = 5  or 6 the system collapses.
On the other hand, the two simulations with initial
values of $\xi_{max}$ = 3 or 4 after a phase of compression re-expand
and within several dynamical timescales achieve a new equilibrium
state that is consistent with their increased total mass. As an example
of a stable merger,
Fig. 5 shows snapshots for a $\xi_{max}$ = 3 main cloud.
After a phase of oscillations, the system settles into a new hydrostatic
equilibrium state. Note that the clouds in this simulation are larger than
in the standard model (Fig. 2) due to the reduced external pressure.

\begin{figure*}[ht]
\begin{center}
\includegraphics[width=0.80\textwidth]{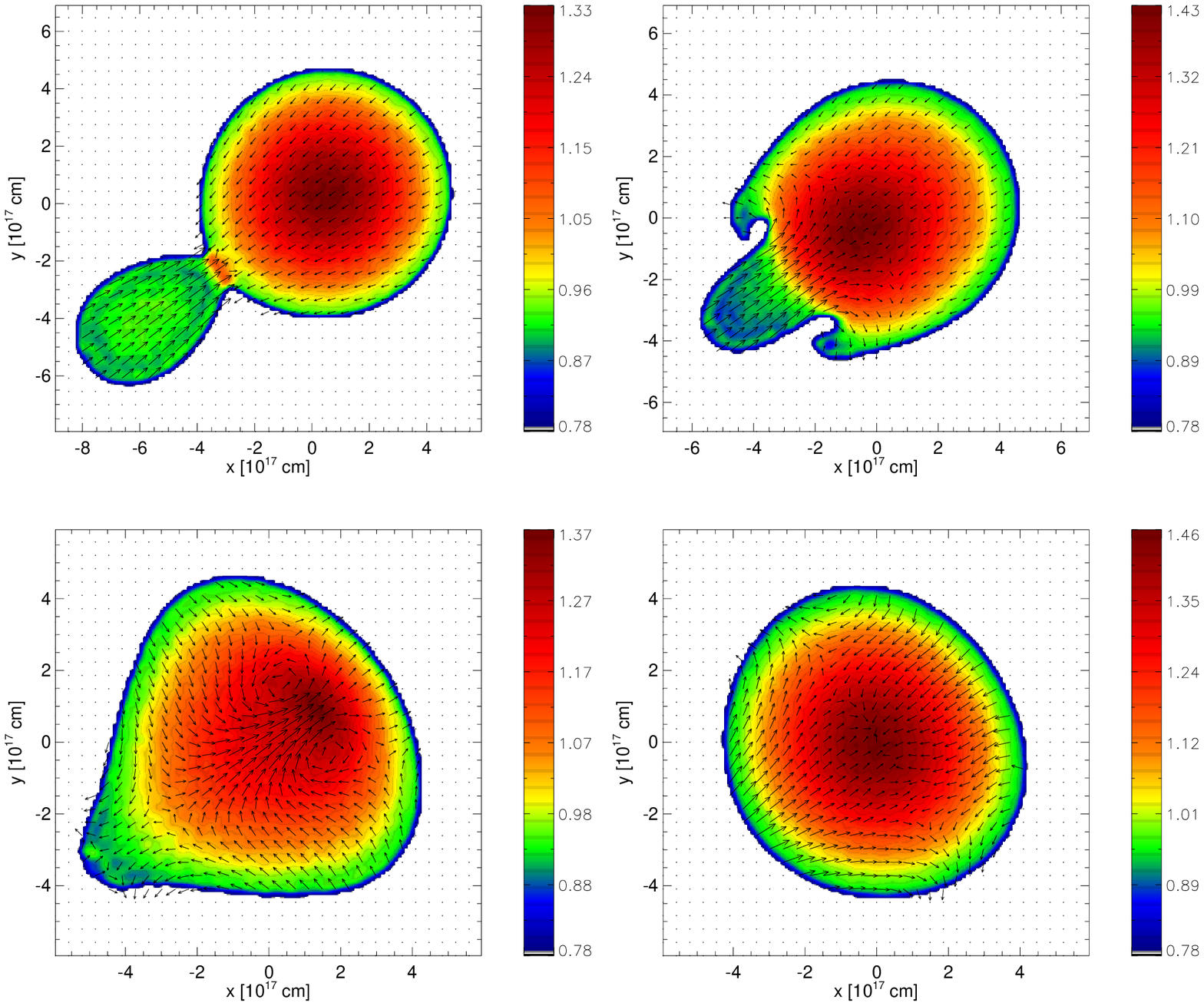}
\end{center}
\caption{
\label{fig5}
Same as Fig. 2, however now with a main cloud
  that is characterised by $\xi_{max}$ = 3. The upper left, upper
  right, lower left and lower right panels show the state of the
  system at evolutionary times of 0.9 million years, 1.7 million
  years, 3.5 million years and 17 million years, respectively.  The
  maximum velocities are 0.16 km/s, 0.25 km/s, 0.017 km/s and 0.013
  km/s, respectively. Note that, in contrast to the unstable merger
  case, here the maximum velocity decreases in the late phases when
  the system goes through a period of damped oscillations and then
  settles into a new pressure confined equilibrium state.}
\end{figure*}

Could Barnard 68 have started with a value lower than $\xi_{max} \leq
4.4$, that is could it survive the collision as a stable, starless
core? The observations provide information not only on the density
structure, characterized by $\xi_{max}$ but also on the phase of
merging, that is the evolutionary time.  The black point in Fig. 4
shows the location of Barnard 68 in the $\xi_{max}$ versus time
diagram. The evolutionary time is no free parameter as it is determined by comparing the
relative positions of both cores in the simulations with the
observations. Excellent agreement exists for initial conditions that
correspond to $\xi_{max} \geq 5$ and that lead to gravitational
collapse.  Interestingly, the situation is different for smaller values.
Although the merger simulation with $\xi_{max} = 4$ 
temporarily leads to a distorted density profile with values of
$\xi_{max} = 7$ as observed for Barnard 68, these values are reached in
a rather late phase of merging after the bullet has disappeared within
the main cloud. We therefore
conclude that Barnard 68 is in fact experiencing a fatal merger 
that leads to gravitational collapse.

\section{Conclusions}

We are then in the fortunate situation of witnessing the collapse of a
Bok globule and the formation of a star like our Sun, or a low-mass
multiple stellar system, in the relative nearby solar
neighborhood. Given the distance to this cloud this would make Barnard
one of the nearest star forming clouds to Earth, a perfect laboratory
to investigate the early phases of gravitational cloud collapse. Barnard 68 has
probably been stable for several million years.
However its current fatal impact with an object of 10\%
its mass has made this globule gravitationally unstable which is
revealed by a dimensionless structural parameter $\xi_{max} = 6.9 \pm
0.2$ that already exceeds the critical value expected for a stable
cloud.  The impact is also mixing gas from the chemically
less evolved outer parts of the main cloud as well as the CO-rich gas
of the bullet with the CO depleted gas in the center of Barnard 68.

Within the next $10^5$ years the bullet will be completely engulfed by
Barnard 68 which at the same time will develop a centrally peaked
surface density profile (dashed curve in Fig. 3) and a velocity field
consistent with large-scale gravitational collapse. $10^5$ years later
a new infrared source will appear in its central region.  An isolated
star like our Sun will be born in our immediate Galactic
neighborhood, probably surrounded by a residual dusty
accretion disk where planets might start forming.

\acknowledgements
We thank Charlie Lada for valuable comments.
The numerical simulations were performed on the local SGI-ALTIX 3700 Bx2, which
was partly funded by the Cluster of Excellence "Origin and Structure of the Universe

\newpage

\begin{center}
APPENDIX
\end{center}

The structure of pressure confined, self-gravitating and isothermal
gas spheres in hydrostatic equilibrium is determined by the
hydrostatic equation

\begin{equation}
\frac{1}{\rho} \vec{\nabla} P = - \vec{\nabla} \Phi
\end{equation}

\noindent where $\rho$ and P is the gas density and the pressure,
respectively.  The gravitational potential $\Phi$ can be calculated 
from the density distribution, using the Poisson equation

\begin{equation}
\nabla^2 \Phi = 4 \pi G \rho
\end{equation}

\noindent with G the gravitational constant. The pressure is given by
the equation of state $P = \rho c_s^2$ where $c_s$ is the
constant isothermal sound speed.

In the case of spherical symmetry, these equations can be combined,
leading to a differential equation for $\Phi$:

\begin{equation}
  \frac{1}{r^2} \frac{d}{dr} \left(r^2 \frac{d \Phi}{dr} \right) = 4 \pi G \rho_c \exp \left(
    - \frac{\Phi (r)}{c_s^2} \right).
\end{equation}

\noindent with

\begin{equation}
\rho (r) = \rho_c \exp \left( - \frac{\Phi (r)}{c_s^2} \right)
\end{equation}

\noindent and $\rho_c = \rho (r=0)$ the central density. We now
introduce the dimensionless variables $\Psi \equiv \Phi/c_s^2$ and
$\xi \equiv \left( 4 \pi G \rho_c / c_s^2 \right)^{0.5}r$.  Equation 5
then reduces to a special form of the Lane-Emden equation

\begin{equation}
\frac{1}{\xi^2} \frac{d}{d \xi} \left( \xi^2 \frac{d \Psi}{d \xi} \right) = \exp (- \Psi ).
\end{equation}

\noindent Given the inner boundary conditions $\Psi (\xi = 0) = 0$ and
$\left( \frac{d \Psi}{d \xi} \right)_{\xi = 0} = 0$, equation 7 can be
integrated numerically. The left panel of Fig. 6 shows the radial
distribution of $\rho/\rho_c = \exp (-\Psi)$ as function of the
dimensional radius $\xi$.

If the gas sphere is confined by an external pressure $P_{ext}$ its
edge with radius R is determined by the condition $\rho (R) = \rho_c
\exp (- \Psi_{max}) = P_{ext}/c_s^2$, that is

\begin{equation}
R = \left( \frac{c_s^2}{4 \pi G \rho_c} \right)^{1/2} \xi_{max}
\end{equation}

\noindent with $\Psi (\xi_{max}) = \Psi_{max} =
-ln \left(\frac{P_{ext}}{\rho_c c_s^2}\right)$.

The total mass of the system is

\begin{eqnarray}
M(r) = 4 \pi \int_{0}^{R} r'^2 \rho (r') dr' = \\
4 \pi \rho_c^{-1/2} 
  \left( \frac{c_s^2}{4 \pi G} \right)^{3/2} \left( \xi^2 \frac{d \Psi}{d \xi} \right)_{\xi = \xi_{max}} \nonumber 
\end{eqnarray}

\noindent We now can define the dimensionless mass

\begin{equation}
  m \equiv \frac{P_{ext}^{1/2} G^{3/2} M}{c_s^4} = \left( 4 \pi \frac{\rho_c}{\rho (R)} \right)^{-1/2}
  \left( \xi^2 \frac{d \Psi}{d \xi} \right)_{\xi = \xi_{max}} .
\end{equation}

\noindent The right panel of Fig. 6 shows $m(\xi_{max})$. m reaches a
maximum $m_{max} = 1.18$ for $\xi_{max}=6.5$, corresponding to a
density contrast of $\rho_c/\rho (R) = 14$.  No equilibrium solution
exists for larger masses. A stability analyses also
shows that for $\xi_{max} > 6.5$ a gaseous configuration is unstable

\begin{figure}[ht]
\begin{center}
\includegraphics[width=0.5\textwidth]{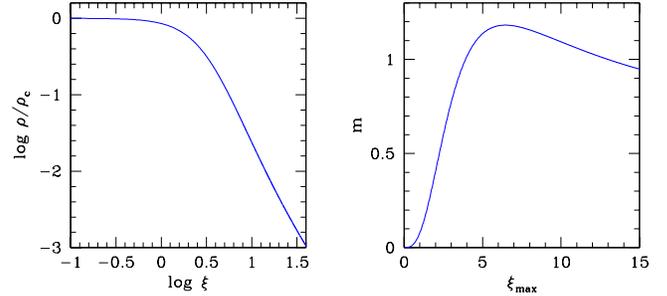}
\end{center}
\caption{
\label{fig6}
The left panel shows the density distribution,
  normalized to the central density $\rho_c$ of an isothermal,
  self-gravitating gas sphere as function of its dimensionless radius
  $\xi$. The dimensionless mass m of a Bonnor-Ebert sphere with radius
  $\xi_{max}$ is shown in the right panel.}
\end{figure}

\end{document}